\documentclass[pra,preprint,preprintnumbers,amsmath,amssymb]{revtex4}

\usepackage{graphicx}
\usepackage{dcolumn}
\usepackage{bm}
\usepackage{amssymb}
\usepackage{amsmath}
\usepackage{amsfonts}
\usepackage[]{graphicx}
\def\beq{\begin{eqnarray}}
\def\eeq{\end{eqnarray}}

\begin{document}
\bibliographystyle{prsty}

\title{ A critical  analysis of `Relative facts do not exist. Relational quantum mechanics is incompatible with quantum mechanics' by Jay  Lawrence, Marcin Markiewicz and Marek \'{Z}ukowski. \footnote{The present arXiv version corrects some typos included in the final version published in  Foundations of Physics [Found. Phys. \textbf{54}:5 (2024)].} }

\begin{abstract}
We discuss a recent work by J.~Lawrence et al.[arxiv.org/abs/2208.11793] criticizing relational quantum mechanics (RQM) and based on a famous nonlocality theorem Going back to Greenberger Horne and Zeilinger (GHZ). Here, we show that the claims presented in this recent work  are unjustified and we debunk the analysis.
\end{abstract}
\author{Aur\'elien Drezet}
\affiliation{Institut N\'eel, UPR 2940, CNRS-Universit\'e Joseph Fourier, 25, rue des Martyrs, 38000 Grenoble, France}

\date{\today}
\maketitle
\section{Introduction}\label{sec1}
\indent The Relational Quantum Mechanics (RQM) is an alternative interpretation of quantum mechanics that was proposed originally by C.~Rovelli~\cite{Rovelli1996,RovelliBook2021,Laudisa,RovelliFP}. RQM can be seen as a logical completion and generalization of the Copenhagen (orthodox) interpretation but where the arbitrariness of Heisenberg's  quantum `shifty-split' or `cut', which  is separating observed and observing subsystems, is taken more seriously. Unlike, the  Copenhagen interpretation  the cut is not confined to the macroscopic domain and  the roles of observed and observing systems are relative and can be inverted. RQM is therefore a more symmetric and general approach.\\
Moreover, recently RQM has been criticized and assessed by various authors. The aim of the present comment (see also the analysis  by Cavalcanti et al. \cite{RovelliReply}) is to give a short reply  to the recent J.~Lawrence et al. article \cite{Zukowski} that concerns RQM and the role of quantum contextuality (for previous claims see \cite{Brukner,Pienaar} and see the replies by Di Biagio and Rovelli \cite{Rovellian} and Drezet \cite{Drezet,Drezetb}). Lawrence et al. have recently replied to the present analysis in \cite{Zukowskireply2}, and to the independent analysis \cite{RovelliReply} in \cite{Zukowskireply1}.\\
\section{The starting point}
\indent Following the recent analysis \cite{Drezet,Drezetb}, I remind  that in RQM the main issue concerns the interpretation of the full wavefunction  $|\Psi_{SO}\rangle$ involving observer (O) and observed system (S).  In RQM the fundamental object  \underline{relatively to (O)} is not   $|\Psi_{SO}\rangle$ but the reduced density matrix
\begin{eqnarray}
\hat{\rho}^{(red.)}_{S|O}=\textrm{Tr}_O[\hat{\rho}_{SO}]=\textrm{Tr}_O[|\Psi_{SO}\rangle \langle \Psi_{SO}|].
\end{eqnarray} 
As it is well known $\hat{\rho}^{(red.)}_{S|O}$ is independent of the basis chosen to represent the degrees of freedom for (O).  In \cite{Drezet} I showed that it  solves the dilemma discussed in \cite{Brukner,Pienaar} concerning the `preferred basis problem'. Here I show that the same features debunk the claims of \cite{Zukowski} concerning non-contextuality.\\
\indent \cite{Zukowski} starts with a GHZ state~\cite{GHZ} for a system S of 3 spins  $m=1,2,3$:
\begin{eqnarray}
|GHZ\rangle_{S}=\frac{1}{\sqrt{2}}[|+1^{(1)},+1^{(1)},+1^{(1)}\rangle_{S_1,S_2,S_3}\nonumber\\+|-1^{(1)},-1^{(1)},-1^{(1)}\rangle_{S_1,S_2,S_3}]
\end{eqnarray}
 where $|p^{(1)}\rangle_{S_m}\equiv |\textrm{sign}(p)z\rangle_{S_m}$ (with $p=\pm 1$) are spin eigenstates along the $z$ direction.
We also have in different spin bases:
\begin{eqnarray}
|GHZ\rangle_S=\frac{1}{2}[|+1^{(2)},+1^{(3)},-1^{(3)}\rangle_{S_1,S_2,S_3}\nonumber\\  +|+1^{(2)},-1^{(3)},+1^{(3)}\rangle_{S_1,S_2,S_3}\nonumber\\    
+|-1^{(2)},+1^{(3)},+1^{(3)}\rangle_{S_1,S_2,S_3}\nonumber\\  +|-1^{(2)},-1^{(3)},-1^{(3)}\rangle_{S_1,S_2,S_3}]
\end{eqnarray} where we used  $|p^{(2)}\rangle_{S_m}=\frac{1}{\sqrt{2}}[|+1^{(1)}\rangle_{S_m} +p^{(2)} |-1^{(1)}\rangle_{S_m}]\equiv |\textrm{sign}(p^{(2)})x\rangle_{S_m}$, and $|p^{(3)}\rangle_{S_m}=\frac{1}{\sqrt{2}}[|+1^{(1)}\rangle_{S_m}+p^{(3)} i|-1^{(3)}\rangle_{S_m}]\equiv |\textrm{sign}(p^{(3)})y\rangle_{S_m}$ (with $p^{(2)},p^{(3)}=\pm 1$). This implies 
\begin{eqnarray}
{\sigma_x}_{S_1}{\sigma_y}_{S_2}{\sigma_y}_{S_3}|GHZ\rangle_{S}=-|GHZ\rangle_{S}
\end{eqnarray}  and 
\begin{eqnarray}
p^{(2)}_{S_1}q^{(3)}_{S_2}r^{(3)}_{S_3}=-1.
\end{eqnarray}
Similar expressions are obtained by circular permutations: 
\begin{eqnarray}
p^{(3)}_{S_1}q^{(2)}_{S_2}r^{(3)}_{S_3}=-1,
\end{eqnarray}
and \begin{eqnarray}
p^{(3)}_{S_1}q^{(3)}_{S_2}r^{(2)}_{S_3}=-1.
\end{eqnarray}
We can also write:
 \begin{eqnarray}
|GHZ\rangle_{S}=\frac{1}{2}[|+1^{(2)},+1^{(2)},+1^{(2)}\rangle_{S_1,S_2,S_3}\nonumber\\  +|+1^{(2)},-1^{(2)},-1^{(2)}\rangle_{S_1,S_2,S_3}\nonumber\\ +|-1^{(2)},+1^{(2)},-1^{(2)}\rangle_{S_1,S_2,S_3}\nonumber\\  +|-1^{(2)},-1^{(2)},+1^{(2)}\rangle_{S_1,S_2,S_3}]
\end{eqnarray} implying 
\begin{eqnarray}
{\sigma_x}_{S_1}{\sigma_x}_{S_2}{\sigma_x}_{S_3}|GHZ\rangle_{S}=+|GHZ\rangle_{S}
\end{eqnarray} and thus
\begin{eqnarray}
p^{(2)}_{S_1}q^{(2)}_{S_2}r^{(2)}_{S_3}=+1.
\end{eqnarray}
 It is well known that quantum mechanics is highly contextual. If the results of spin measurements were non-contextual   Eqs.  5,6,7 and 10 could be true together and this, clearly, is not possible: multiplying  Eqs. 5,6,7 contradicts Eq. 10. This incompatibility also leads to a well known proof of quantum nonlocality without inequality~\cite{GHZ}.\\
 \section{Debunking a paradox}
 \indent In \cite{Zukowski} the authors use the previous results. First, they consider a single observer Alice (A) (composed of 3 qubits $A_1,A_2,A_3$) who measures the spins  of the GHZ system S in the  y bases. The GHZ states of Eq.~2  reads in the y bases: 
 \begin{eqnarray}
|GHZ\rangle_{S}=\sum_{p,q,r}C^{333}_{pqr}|p^{(3)},q^{(3)},r^{(3)}\rangle_{S_1,S_2,S_3}
\end{eqnarray}
 where the amplitudes $C^{333}_{pqr}$ are non vanishing for each of the 8 combinations $p,q,r$  and where $|C^{333}_{pqr}|^2=\frac{1}{8}$. After entanglement with Alice's qubits we have   
\begin{eqnarray}
|GHZ\rangle_{SA}=\sum_{p,q,r}C^{333}_{pqr}|p^{(3)}\rangle_{SA_1}|q^{(3)}\rangle_{SA_2}|r^{(3)}\rangle_{SA_3}
\end{eqnarray}
 with $|k^{(3)}\rangle_{SA_m}:=|k^{(3)}\rangle_{S_m}|k^{(3)}\rangle_{A_m}$ and $|k^{(3)}\rangle_{A_m}$ is the state of the qubit $A_m$, $m=1,2,3$ (for $k=\pm1$).\\
 \indent Importantly, for Alice this state shows no correlation.  This is clear in RQM where the reduced density matrix reads $\hat{\rho}^{(red.)}_{S|A}=\textrm{Tr}_A[\hat{\rho}_{SA}]$:
\begin{eqnarray}
\hat{\rho}^{(red.)}_{S|A}=\frac{1}{8}\sum_{p,q,r}|p^{(3)}\rangle_{S_1}{}_{S_1}\langle p^{(3)}|\otimes|q^{(3)}\rangle_{S_2}{}_{S_2}\langle q^{(3)}|\nonumber\\ \otimes|r^{(3)}\rangle_{S_3}{}_{S_3}\langle r^{(3)}|
\end{eqnarray}  
Due to decoherence i.e., entanglement with the environement (Alice)  we have lost coherence and correlations between spins.   In particular we have
\begin{eqnarray}
\textrm{Tr}_S[{\sigma_x}_{S_1}{\sigma_x}_{S_2}{\sigma_x}_{S_3}\hat{\rho}^{(red.)}_{S|A}]=0,\nonumber\\
\textrm{Tr}_S[{\sigma_x}_{S_1}{\sigma_y}_{S_2}{\sigma_y}_{S_3}\hat{\rho}^{(red.)}_{S|A}]=0,\nonumber\\
\textrm{Tr}_S[{\sigma_y}_{S_1}{\sigma_x}_{S_2}{\sigma_y}_{S_3}\hat{\rho}^{(red.)}_{S|A}]=0,\nonumber\\
\textrm{Tr}_S[{\sigma_y}_{S_1}{\sigma_y}_{S_2}{\sigma_x}_{S_3}\hat{\rho}^{(red.)}_{S|A}]=0,\nonumber\\
\end{eqnarray} which contrast with 
\begin{eqnarray}
\textrm{Tr}_S[{\sigma_x}_{S_1}{\sigma_x}_{S_2}{\sigma_x}_{S_3}\hat{\rho}^{(red.)}_{S}]=+1,\nonumber\\
\textrm{Tr}_S[{\sigma_x}_{S_1}{\sigma_y}_{S_2}{\sigma_y}_{S_3}\hat{\rho}^{(red.)}_{S}]=-1,\nonumber\\
\textrm{Tr}_S[{\sigma_y}_{S_1}{\sigma_x}_{S_2}{\sigma_y}_{S_3}\hat{\rho}^{(red.)}_{S}]=-1,\nonumber\\
\textrm{Tr}_S[{\sigma_y}_{S_1}{\sigma_y}_{S_2}{\sigma_x}_{S_3}\hat{\rho}^{(red.)}_{S}]=-1,\nonumber\\
\end{eqnarray}
 In RQM Eq.~15 actually describes the correlations available to A before the interaction occurred, i.e., when the full state of SA is still factorized. It represents a catalog of knowledge or potentiality in the sense of Heisenberg.   The actualization of measurements in RQM is a debatable issue  and we will not consider this problem here (see e.g.\cite{Drezet}).\\
\indent In the next step the authors of \cite{Zukowski} consider  a second observer Bob (B) (also composed of 3 qubits $B_1,B_2,B_3$) who measures the spins  of the entangled GHZ system SA   in the x bases of the joint system. In analogy with Eq. 8 we write after entanglement with Bob qubits:
 \begin{eqnarray}
|GHZ\rangle_{SAB}=\frac{1}{2}\sum_{p,q,r}|p^{(2)}\rangle_{SAB_1}|q^{(2)}\rangle_{SAB_2}|r^{(2)}\rangle_{SAB_3}
\end{eqnarray}
 with $|k^{(2)}\rangle_{SAB_m}:=|k^{(2)}\rangle_{SA_m}|k^{(2)}\rangle_{B_m}$ and $|k^{(2)}\rangle_{B_m}$ is the state of the qubit $B_m$, $m=1,2,3$ (for $k=\pm1$) and where we introduced the entangled `x' states for the SA system: $|k^{(2)}\rangle_{SA_m}=\frac{1}{\sqrt{2}}[|+1^{(3)}\rangle_{SA_m}+i k^{(2)}|-1^{(3)}\rangle_{SA_m}]\equiv |\textrm{sign}(k^{(2)})x\rangle_{SA_m}$. Crucially,  the  numbers  $p,q,r=\pm1$ in Eq. 16 must obey the GHZ-constraint:
 \begin{eqnarray}
p^{(2)}_{SAB_1}q^{(2)}_{SAB_2}r^{(2)}_{SAB_3}=+1.
\end{eqnarray}
Once more, in RQM we need to consider the reduced density matrix $\hat{\rho}^{(red.)}_{SA|B}=\textrm{Tr}_B[\hat{\rho}_{SAB}]$: 
 \begin{eqnarray}
\hat{\rho}^{(red.)}_{SA|B}= \frac{1}{4}\sum_{p,q,r}|p^{(2)}\rangle_{SA_1}{}_{SA_1}\langle p^{(2)}|\otimes|q^{(2)}\rangle_{SA_2}{}_{SA_2}\langle q^{(2)}|\nonumber\\  \otimes|r^{(2)}\rangle_{SA_3}{}_{SA_3}\langle r^{(2)}|\nonumber\\
\end{eqnarray} with again  $pqr=+1$.  This defines the information available to B in RQM  and this density matrix shows partial coherence  since we have  
\begin{eqnarray}
\textrm{Tr}_{SA}[{\sigma_x}_{SA_1}{\sigma_x}_{SA_2}{\sigma_x}_{SA_3}\hat{\rho}^{(red.)}_{SA|B}]=+1,\nonumber\\
\textrm{Tr}_{SA}[{\sigma_x}_{SA_1}{\sigma_y}_{SA_2}{\sigma_y}_{SA_3}\hat{\rho}^{(red.)}_{SA|B}]=0,\nonumber\\
\textrm{Tr}_{SA}[{\sigma_y}_{SA_1}{\sigma_x}_{SA_2}{\sigma_y}_{SA_3}\hat{\rho}^{(red.)}_{SA|B}]=0,\nonumber\\
\textrm{Tr}_{SA}[{\sigma_y}_{SA_1}{\sigma_y}_{SA_2}{\sigma_x}_{SA_3}\hat{\rho}^{(red.)}_{SA|B}]=0.\nonumber\\
\end{eqnarray} The first line of Eq. 19 is of course reminiscent of Eq. 17 and shows that there is a preferred pointer basis defined by the specific measurement protocol.   This is associated with a specific interaction Hamiltonian $H_{SA,B}$ leading to the state given by Eq. 16.\\
\indent  However, the authors of \cite{Zukowski} didn't correctly analyze the structure of RQM and the meaning of relative facts. They claim that   we can find relations between facts or information available to Bob and facts or information available to Alice.  This we show below is actually a misunderstanding. More precisely, they  consider that Bob only measures one of the 3 qubits belonging to SA. In the following we consider the particular case $m=1$ and  therefore only the system $SA_1$ will interact with $B_1$. This requires to let the two other qubits of Bob $B_2$ and $B_3$  in their respective ground states. As the authors show we get the new state   
  \begin{eqnarray}
|GHZ'\rangle_{SAB}=\frac{1}{2}\sum_{p,q,r}|p^{(2)}\rangle_{SAB_1}|q^{(3)}\rangle_{SA_2}|in\rangle_{B_2}|r^{(3)}\rangle_{SA_3}|in\rangle_{B_3}
\end{eqnarray} with now the constraint: 
\begin{eqnarray}
p^{(2)}_{SAB_1}q^{(3)}_{SA_2}r^{(3)}_{SA_3}=-1.
\end{eqnarray}
Of course, we could develop two similar procedures acting only on the qubit $B_2$ or alternatively the qubit $B_3$. We will obtain two different  states  $|GHZ''\rangle_{SAB}$ and $|GHZ'''\rangle_{SAB}$ leading to the relations
\begin{eqnarray}
p^{(3)}_{SA_1}q^{(2)}_{SAB_2}r^{(3)}_{SA_3}=-1,
\end{eqnarray} and 
\begin{eqnarray}
p^{(3)}_{SA_1}q^{(3)}_{SA_2}r^{(2)}_{SAB_3}=-1.
\end{eqnarray}
With these mathematical properties the deduction of \cite{Zukowski} goes as follows:\\
\indent  i) It is visible from Eq. 21 (and similarly for Eqs. 22,23 by cyclic permutation) that  the number $p^{(2)}_{SAB_1}$ characterizes the entangled system $SAB_1$ whereas $q^{(3)}_{SA_2}$ and $r^{(3)}_{SA_3}$ characterize $SA_2$ and $SA_3$. Therefore it is tempting to call $p^{(2)}_{SAB_1}$ a relative fact for Bob and $q^{(3)}_{SA_2}$ and $r^{(3)}_{SA_3}$ relative facts for Alice. This idea, which is central for their paper, is clearly summarized by the analysis surrounding their equation 17 in  \cite{Zukowski}. They call $\mathcal{B}_m$ the number  $k^{(2)}_{SAB_m}$ defined in our Eqs. 17, 21-23 and similarly they call $\mathcal{A}_m$ the number  $k^{(3)}_{SA_m}$. Assuming this we go to the next step of their `no-go' deduction.\\
\indent  ii) The 4 relations Eq. 17, 21-23.  are clearly incompatible.  If we multiply   Eq. 21 by Eqs. 22 and 23 we obtain
 \begin{eqnarray}
p^{(2)}_{SAB_1}q^{(2)}_{SAB_2}r^{(2)}_{SAB_3}(p^{(3)}_{SA_1}q^{(3)}_{SA_2}r^{(3)}_{SA_3})^2\nonumber\\=p^{(2)}_{SAB_1}q^{(2)}_{SAB_2}r^{(2)}_{SAB_3}=-1,
\end{eqnarray}
  which clearly contradicts Eq. 17. Now as they emphasize it clearly: `\textit{Note that, most importantly for the sequel, the
three [unitary] transformations [acting on  $m = 1, 2, 3$] mutually commute, and thus their order of application is immaterial.}'. In other words: Since the 3 operations leading to $|GHZ'\rangle_{SAB}$, $|GHZ''\rangle_{SAB}$,$|GHZ''\rangle_{SAB}$ and thus Eqs. 21-23 are acting `locally' only on one of the sub systems  $SA_m$ their meaning should be non contextual and absolute.  This following \cite{Zukowski} justifies why we should apriori compare these states with  $|GHZ\rangle_{SAB}$ and Eq. 17. This noncontextual reading is what they believe is contained in RQM.\\
\indent If we accept this reasoning then RQM contradicts quantum mechanics. Relative facts for Alice and Bob are contradictory.\\  
\indent However, this deduction is false and we now debunk the contradiction. First, consider  point i): It is clear that $p^{(2)}_{SAB_1}$ must be a relative fact for Bob. However, there is no reason to consider that $q^{(3)}_{SA_2}$ and $r^{(3)}_{SA_3}$ should be relative facts for Alice. Actually, the quantum state Eq. 20 contains quantum numbers $p,q,r$ but in RQM the fundamental description for Bob is the reduced  density matrix $\hat{\rho'}^{(red.)}_{SA|B}=\textrm{Tr}_B[\hat{\rho'}_{SAB}]$ obtained by using $\hat{\rho'}_{SAB}=|GHZ'\rangle_{SAB}{}_{SAB}\langle|GHZ'|$. We have: 
\begin{eqnarray}
\hat{\rho'}^{(red.)}_{SA|B}=\frac{1}{2}|+1^{(2)}\rangle_{SA_1}{}_{SA_1}\langle +1^{(2)}|\nonumber\\\otimes|\Phi\rangle_{SA_2,SA_3}{}_{SA_2,SA_3}\langle \Phi| \nonumber\\
+\frac{1}{2}|-1^{(2)}\rangle_{SA_1}{}_{SA_1}\langle -1^{(2)}|\nonumber\\\otimes|\Psi\rangle_{SA_2,SA_3}{}_{SA_2,SA_3}\langle \Psi|
\end{eqnarray}  
where 
\begin{eqnarray}
|\Phi\rangle_{SA_2,SA_3}=\frac{1}{\sqrt{2}}[|+1^{(3)}\rangle_{SA_2}|-1^{(3)}\rangle_{SA_3}\nonumber\\+|-1^{(3)}\rangle_{SA_2}|+1^{(3)}\rangle_{SA_3}],\nonumber\\
|\Psi\rangle_{SA_2,SA_3}=\frac{1}{\sqrt{2}}[|+1^{(3)}\rangle_{SA_2}|+1^{(3)}\rangle_{SA_3}\nonumber\\+|-1^{(3)}\rangle_{SA_2}|-1^{(3)}\rangle_{SA_3}]
\end{eqnarray} 
are two Bell states. The fact that EPR-like Bell states are present show that there is a coherence preserved in the description  of the system  SA by Bob. In particular we deduce
\begin{eqnarray}
\textrm{Tr}_{SA}[{\sigma_x}_{SA_1}{\sigma_x}_{SA_2}{\sigma_x}_{SA_3}\hat{\rho'}^{(red.)}_{SA|B}]=+1,\nonumber\\
\textrm{Tr}_{SA}[{\sigma_x}_{SA_1}{\sigma_y}_{SA_2}{\sigma_y}_{SA_3}\hat{\rho'}^{(red.)}_{SA|B}]=-1,\nonumber\\
\textrm{Tr}_{SA}[{\sigma_y}_{SA_1}{\sigma_x}_{SA_2}{\sigma_y}_{SA_3}\hat{\rho'}^{(red.)}_{SA|B}]=0,\nonumber\\
\textrm{Tr}_{SA}[{\sigma_y}_{SA_1}{\sigma_y}_{SA_2}{\sigma_x}_{SA_3}\hat{\rho'}^{(red.)}_{SA|B}]=0.\nonumber\\
\end{eqnarray} 
 The second line is of course reminiscent of Eq.~21 but the meaning is here different from what is claimed in \cite{Zukowski}. Indeed, this a relation  for Bob measurements not for Alice.  The idea to have a mixture of information for Bob and Alice in the same Eq.~21 is thus unjustified and is not a part of RQM but instead of the reading of RQM made by the authors of \cite{Zukowski}. The first line is also very interesting: It shows coherence associated with the  $|\Phi\rangle_{SA_2,SA_3}$ and $|\Psi\rangle_{SA_2,SA_3}$  Bell states. Moreover this result can be compared with Eq.~19 for $\hat{\rho}^{(red.)}_{SA|B}$. The difference clearly stresses that the measurement  procedures are more invasive that claimed in \cite{Zukowski}.  In RQM like in the Copenhagen interpretation   the experimental context   and the choice of the interaction Hamiltonian is key.    This allows us us to answer to point ii) concerning non-contextuality.  Indeed,   the non contextuality supposed by the authors of \cite{Zukowski} is not a part of the RQM. For RQM (like in the orthodox interpretation) the different contexts are not compatible and we have no right to compare Eqs.~17 and Eqs.~21-23 as claimed.  This would  otherwise contradict the central axioms of RQM and therefore the reasoning discussed in \cite{Zukowski} is unjustified.\\
 \indent We conclude this analysis  by adding that RQM is a self-consistent interpretation of quantum mechanics extending the old Copenhagen interpretation. The formalism is perfectly agreeing with standard quantum mechanics and recovers the orthodox interpretation at the limit were observers are essentially macroscopic but without the problems concerning the definition of agents in Qbism.   Several issues concerning interactions can still be debated but we should not focus on wrong problems.  Often objections are done without carefully considering the philosophy of RQM but by adding some prejudices or preconceptions concerning the interpretation (i.e., \cite{Zukowski,Brukner,Pienaar}). I hope this note will contribute to clarify a bit this issue. \\
 \indent The author thanks very useful discussions with Carlo Rovelli, Eric Cavalcanti,  and Marek \'{Z}ukowski.
          
\section*{Competing Interest}
The Author declares no competing interest for this work. 

 \section*{Data Availability Statement}
Data Availability Statement: No Data associated in the manuscript.

\end{document}